\documentclass[11pt,a4paper,final]{iopart}
\usepackage{graphicx}
\usepackage[breaklinks=true,colorlinks=true,linkcolor=blue,urlcolor=blue,citecolor=blue]{hyperref}
\expandafter\let\csname equation*\endcsname\relax
\expandafter\let\csname endequation*\endcsname\relax
\usepackage{amsmath}
\usepackage{amssymb}
\usepackage{amsfonts}
\usepackage{physics}
\usepackage{braket}
\usepackage[dvipsnames]{xcolor}
\usepackage[T1]{fontenc}

\newcommand\undermat[2]{
  \makebox[0pt][l]{$\smash{\underbrace{\phantom{%
    \begin{matrix}#2\end{matrix}}}_{\text{$#1$}}}$}#2}

\usepackage[normalem]{ulem}

\begin{document}

\title[]{Spin-orbit coupling in the presence of strong atomic correlations}

\author{Ayaka Usui$^{1}$, Thom\'{a}s Fogarty$^{1}$, Steve Campbell$^{2}$, Simon A. Gardiner$^{3}$, and Thomas Busch$^{1}$}
\address{$^1$Quantum Systems Unit, Okinawa Institute of Science and Technology Graduate University, Okinawa, Japan}
\address{$^2$University College Dublin, Belfield, Dublin 4, Ireland}
\address{$^3$Joint Quantum Centre (JQC) Durham-Newcastle, Department of Physics, Durham University, Durham DH1 3LE, United Kingdom}
\ead{\mailto{ayaka.usui@oist.jp}}

\begin{abstract}
We explore the influence of contact interactions on a synthetically spin-orbit coupled system of two ultracold trapped atoms. Even though the system we consider is bosonic, we show that a regime exists in which the competition between the contact and spin-orbit interactions results in the emergence of a ground state that contains a significant contribution from the anti-symmetric spin state. This ground state is unique to few-particle systems and does not exist in the mean-field regime. The transition to this state is signalled by an inversion in the average momentum from being dominated by centre-of-mass momentum to relative momentum and also affects the global entanglement shared between the real- and pseudo-spin spaces. Indeed, competition between the interactions can also result in avoided crossings in the ground state which further enhances these correlations. However, we find that correlations shared between the pseudo-spin states are strongly depressed due to the spin-orbit coupling and therefore the system does not contain spin-spin entanglement. 
\end{abstract}
\vspace{2pc}

\section{Introduction\label{sec:intro}}

Spin-orbit coupling (SOC) is an effect that was initially discussed in systems of charged particles. It is of large prominence in condensed matter physics and underlies, for example, the appearance of the spin Hall effect \cite{Nagaosa2010} or of topological insulator states \cite{Hasan2010}. However, as controlling the SOC parameters in condensed matter systems is usually hard, exploring all possible states and limits is often not feasible. The recent progress in implementing synthetic SOC in systems of cold neutral atoms has led to significant progress in this respect and controllable systems with long coherence times and a lack of impurities are now experimentally available. In particular, Bose-Einstein condensates (BECs) coupled by Raman lasers can be used to generate synthetic SOC, using a pseudo-spin realised by two internal states of the atoms and selective momentum transfer. In recent years SOC has been realised in (pseudo) spin-1/2 Bose gases \cite{Lin2011a,Zhang2012a}, spin-1 Bose gases \cite{Campbell2016b} and also in Fermi gases \cite{Wang2012,Cheuk2012}.

The effect SOC has in BECs can be understood within the mean-field approximation of a two-component gas by calculating the dispersion relation and the related phase diagram \cite{Li2012,Zhang2012b,Huang2015,Zhang2016}. In this regime three distinct phases can exist when the two components are in the miscible regime and when all interactions are repulsive. For zero or small Raman coupling, the system exhibits a striped density pattern, which originates from a superposition of states with positive and negative momentum. In this parameter regime the gas is therefore in a supersolid phase. If the Raman coupling strength is increased, the system enters the magnetised phase, where the ground state is degenerate between the positive and the negative-momentum state. Finally, increasing the Raman coupling strength even further, the system enters the single minimum phase, where the ground state no longer carries momentum. However, the mean-field approximation imposes classical fields and ignores the quantum fluctuations, which means that certain effects can be lost. 

While solving large many-body systems with approaches beyond mean-field is a very difficult task and only possible in special cases \cite{Lieb1963a,Lieb1963b,Sutherland2004}, few-particle systems can actually be amenable to exact treatments across the whole range of interactions and correlation strengths \cite{Busch1998,Zinner_2014,zinner2017,Bougas2019,Sowinski2019}. Several treatments of SOC in such systems have already been carried out  \cite{Guan2014,Guan2015,Schillaci2015,Zhu2016,Blume2018,Mujal2019,Guan2019} and, for example, a mapping to an effective spin model was recently suggested by a perturbative approach to account for weak Raman coupling \cite{Guan2015}. It was also shown that, while there is no entanglement in the mean-field regime, in two particle systems the ground state can be maximally entangled in the pseudo-spin space \cite{Zhu2016}. 

In this work we exactly solve the system of two interacting particles in a harmonic trap in the presence of SOC, using a numerical approach that allows us to obtain accurate solutions for any strength of the contact and SOC interactions. We find that the interplay between the contact interactions and the SOC leads to lifting of degeneracies in the energy spectrum, which in certain parameter regimes results in the appearance of a unique ground state that is not revealed within a mean-field treatment. Our choice of basis provides a convenient means to describe the composition of this ground state, and we show that it consists of a finite component of the anti-symmetric pseudo-spin state. We therefore refer to this state as the anti-symmetric (AS) ground state. This AS ground state is distinct from the three phases existing in the mean-field limit and can be signalled by the non-classical correlations between the real space and the pseudo-spin degrees of freedom. Our results are a useful contribution to the understanding of the emergent behaviour of quantum systems and the bridging of the gap between single and many-body states. The framework presented is also well suited to investigate these systems' dynamical properties \cite{GarciaMarch2016,Koutentakis2019,Barfknecht2019}.

The manuscript is organised as follows. In Sec.~\ref{sec:hamiltonian} we introduce the Hamiltonian describing the two-particle system in the presence of SOC in position space and expand the atomic position Hilbert space of the system in harmonic oscillator basis states associated with the centre of mass and relative motional degrees of freedom, using this to embed the bosonic symmetry of the system. In Sec.~\ref{sec:results} we investigate the effects of the interactions and SOC on the energy spectrum and the ground states, and in particular discuss the appearance of the AS ground state. We also explore the entanglement in the pseudo-spin degrees of freedom and between the real space components and the pseudo-spin components. In Sec.~\ref{sec:conclusion} we conclude. Details about the systematic representation of the Hamiltonian in matrix form are given in \nameref{app:matH}.

\section{\label{sec:hamiltonian}Formalism}
We consider an effective one-dimensional model of two repulsively interacting bosons in a harmonic trap in the presence of SOC. The pseudo-spin is given by two hyper-fine states of each atom and the coupling can be experimentally realised by using a two photon process and interpreting it as an effective Raman coupling with an additional momentum boost \cite{Lin2011a}. The Hamiltonian for such a system has the form
\begin{equation} \label{eq:hamiltonian}
    H = \sum_{j=1}^{2} \left[ \frac{p_{j}^2}{2m}+\frac{1}{2}m\omega^2 x_j^2 + \frac{\hbar k_{\mathrm{soc}}}{m}p_{j}\sigma_{z}^{(j)} + \frac{\hbar\Omega}{2}\sigma_{x}^{(j)} + \frac{\hbar\Delta}{2}\sigma_z^{(j)}\right] 
    + H_{\mathrm{int}}(|x_1-x_2|)\;,
\end{equation}
where $m$ is the mass of the particles, $\omega$ is the trap frequency, $k_{\mathrm{soc}}$ is the SOC strength, $\Omega$ is the Raman coupling strength, and $\Delta$ is a detuning. The $\sigma_i$ are the Pauli matrices. The third term describes SOC which couples the momentum and pseudo-spin degrees of freedom. Experimentally, this coupling is facilitated by the momentum difference between the two Raman laser pulses \cite{Lin2011a} and $k_{\mathrm{soc}}$ is the projected wave number determined by the wavelength and the angle of intersection between the Raman lasers \cite{Hamnerthesis}. At low temperatures one can assume that the scattering between the particles has s-wave character only, which permits description of the interaction potentials by point-like $\delta$-functions with internal state-dependent coupling constants
\begin{align}
\begin{split}
    \expval{H_{\mathrm{int}}}{x_1,\uparrow;x_2, \uparrow} &= g_{\uparrow\uparrow} \delta(x_1-x_2)\;, \\
    \expval{H_{\mathrm{int}}}{x_1,\uparrow;x_2,\downarrow} &= g_{\uparrow\downarrow}\delta(x_1-x_2) = \expval{H_{\mathrm{int}}}{x_1,\downarrow;x_2,\uparrow} \;,\\
    \expval{H_{\mathrm{int}}}{x_1,\downarrow;x_2,\downarrow} &= g_{\downarrow\downarrow} \delta(x_1-x_2)\;.
\end{split}
\end{align}
These coupling constants are a function of the respective 3D scattering lengths, $a_{3\mathrm{D}}$, via $g_{i,j}=4\hbar^{2}a_{3\mathrm{D}}/\left(1-C a_{3\mathrm{D}}/d_{\perp}\right)md_{\perp}^{2}$ for $i,j=\downarrow,\uparrow$, where $d_{\perp}=\sqrt{\hbar/m\omega_{\perp}}$ quantifies the trap width in the transverse direction for a trap of frequency $\omega_{\perp}$ and the constant $C$ is given by $C\approx 1.4603$ \cite{Olshanii1998}. For simplicity and to discuss the interesting physics clearly, we neglect the effects of the detuning by setting $\Delta=0$ and restrict ourselves to the symmetric situation by assuming that the interactions between atoms in the same spin state are independent of the spin direction, $g_{\uparrow\uparrow}=g_{\downarrow\downarrow}=g$. A generalisation to arbitrary values for $g_{\uparrow\uparrow}$ and $g_{\downarrow\downarrow}$ is technically straightforward. In the absence of Raman coupling, the Hamiltonian \eqref{eq:hamiltonian} is diagonal within the pseudo-spin basis, and the solutions are given by the eigenstates of the bare harmonic oscillator Hamiltonian for two interacting particles with an added momentum boost of $\sqrt{2}\hbar k_{\mathrm{soc}}$ due to the SOC \cite{Guan2014}. In the presence of Raman coupling, the Hamiltonian \eqref{eq:hamiltonian} has off-diagonal terms that couple the different pseudo-spin basis states.

To clarify the symmetries inherent in the system, let us introduce scaled centre-of-mass (COM) and relative coordinates as $x_\pm=(x_1 \pm x_2)/\sqrt{2}$ and an alternative pseudo-spin basis given by $\ket{\downarrow\downarrow}$, $\ket{\uparrow\uparrow}$,  $\ket{\mathrm{S}}=(\ket{\downarrow\uparrow}+\ket{\uparrow\downarrow})/\sqrt{2}$ and $\ket{\mathrm{A}}=(\ket{\downarrow\uparrow}-\ket{\uparrow\downarrow})/\sqrt{2}$. The first three states of this basis are symmetric under exchange of spins, and the last is anti-symmetric. In this basis the Hamiltonian can be written as
\begin{equation} \label{eq:Hoperator}
\begin{split}
    H 
    &= 
    \hbar\omega\left(\hat{a}^{\dagger}_{+}\hat{a}_{+}+\frac{1}{2}\right)
    +\hbar\omega\left(\hat{a}^{\dagger}_{-}\hat{a}_{-}+\frac{1}{2}\right) \\
    &\quad-i\Lambda\left(\hat{a}^{\dagger}_{+} - \hat{a}_{+}\right)
    \left(\dyad{\downarrow\downarrow}-\dyad{\uparrow\uparrow}\right)
    -i\Lambda\left(\hat{a}^{\dagger}_{-} - \hat{a}_{-}\right)
    \left(\dyad{\mathrm{S}}{\mathrm{A}}+\dyad{\mathrm{A}}{\mathrm{S}}\right) \\
    &\quad+\Upsilon\left(\dyad{\mathrm{S}}{\downarrow\downarrow}+\dyad{\downarrow\downarrow}{\mathrm{S}}+\dyad{\mathrm{S}}{\uparrow\uparrow}+\dyad{\uparrow\uparrow}{\mathrm{S}}\right) + H_{\mathrm{int}}\;,
\end{split}
\end{equation} 
where the $\hat{a}^{\dagger}_{\pm}, \hat{a}_{\pm}$ are creation and annihilation operators for modes in the COM and relative coordinate space, $\Lambda=\hbar k_{\mathrm{soc}}\sqrt{\hbar\omega/m}$ and $\Upsilon=\hbar\Omega/\sqrt{2}$. The basis states of this Hamiltonian are labelled as $\ket{n_+,n_-,\eta}$ for the quantum numbers $n_+, n_-$ of the COM and relative motion, and the pseudo-spin states given by $\eta\in\lbrace \downarrow\downarrow,\mathrm{S},\uparrow\uparrow,\mathrm{A} \rbrace$. Since the system is bosonic, all the states are symmetric with respect to particle exchange. While particle exchange does not affect the COM coordinate, the same is not true for the relative motion. Thus, we impose a restriction on the quantum number $n_-$ of the relative motion as
\begin{equation}
    n_- =
    \begin{cases}
    2u & \text{for} \ \  \eta\in\left\{\downarrow\downarrow, \mathrm{S}, \uparrow\uparrow\right\} \\
    2u+1 & \text{for} \ \ \eta\in\left\{\mathrm{A}\right\}
    \end{cases}\;,
\end{equation}
for integer $u\geq 0$. Thus, the basis states are given by
\begin{equation} \label{eq:basis}
    \ket{n,2u,\downarrow\downarrow},
    \ket{n,2u,\mathrm{S}},
    \ket{n,2u,\uparrow\uparrow},
    \ket{n,2u+1,\mathrm{A}}\;,
\end{equation}
for integer $n\geq0$ and the eigenstates of the Hamiltonian \eqref{eq:Hoperator} can be written as
\begin{equation}
    \psi_{j} = \sum_{n_+,n_-,\eta} a^{(j)}_{n_+,n_-,\eta} \ket{n_+,n_-,\eta}
\end{equation}
for $\{n_+,n_-,\eta\}\in\{n,2u,\downarrow\downarrow\},\{n,2u,\mathrm{S}\},\{n,2u,\uparrow\uparrow\},\{n,2u+1,\mathrm{A}\}$. The interaction part of the Hamiltonian can be expanded within this basis and described using the eigenstates $\phi_n(x)$ of the harmonic oscillator by
\begin{equation} \label{eq:Hint}
\begin{split}
    \mel{n_+,n_-,\eta}{H_{\mathrm{int}}}{n'_+,n'_-,\eta'}
    &=
    g_{\eta}
    \delta_{n_+,n'_+}\delta_{\eta,\eta'}
    \int dx_- \delta(x_1-x_2)\phi_{n_-}(x_-)\phi_{n'_-}(x_-) \\
    &=
    \frac{g_{\eta}}{\sqrt{2}}
    \delta_{n_+,n'_+}\delta_{\eta,\eta'}\phi_{n_-}(0)\phi_{n'_-}(0)\;,
\end{split}
\end{equation}
which follows from Eq.~\eqref{eq:Hintapp} together with the representation given in Eq.~\eqref{eq:fuv} and with $g_{\mathrm{S}}=g_{\mathrm{A}}=g_{\uparrow\downarrow}$. Note that the states with $\eta=\mathrm{A}$ do not feel the contact interaction because $\phi_{n_{-}}(0)=0$ for $n_-$ odd and that the interactions in general lead to energy shift and coupling between different basis states and energy shift. See the \nameref{app:matH} for more details on how to systematically construct a matrix representation of this Hamiltonian.

\begin{figure}[tbp]
    \includegraphics[width=\textwidth]{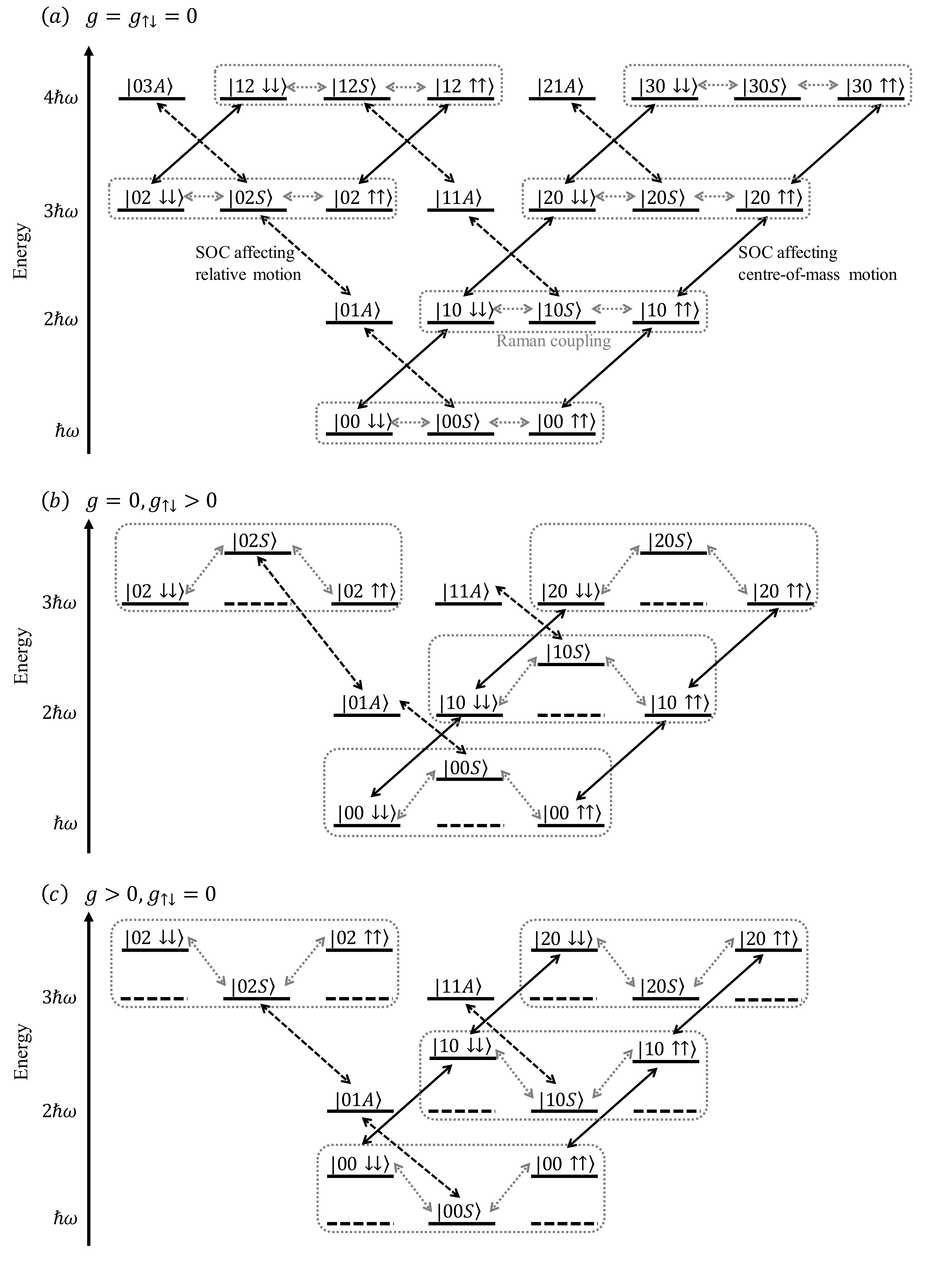}
    \caption{Energy level diagrams for Hamiltonian \eqref{eq:Hoperator}, using as a basis the eigenstates in the absence of SOC and Raman coupling. The basis states are labelled as $\ket{n_+,n_-,\eta}$ for the quantum numbers $n_+, n_-$ of the COM and relative motion, and the pseudo-spin states given by $\eta\in\lbrace \downarrow\downarrow,\mathrm{S},\uparrow\uparrow,\mathrm{A} \rbrace$. Black arrows represent transitions due to SOC, with full arrows exciting COM motion and dashed arrows exciting relative motion. Gray, dotted arrows represent transitions due to Raman coupling. In panel (a) the interaction effects are absent, whereas in panels (b) and (c) the level shifts due to the interactions can be seen.}
    \label{fig:diag_energylevel}
\end{figure}

This description allows for a clear and intuitive interpretation of the basis states and the coupling between them. For simplicity let us first consider non-interacting particles, for which the basis states can be grouped according to harmonic oscillator energies [rows in Fig.~\ref{fig:diag_energylevel}(a)]. The SOC terms then lead to two possible transitions, depending on the spin-states of the atoms. If both particles have the same spin, states with different COM quantum numbers, $n_+$, and the same relative-motion quantum number, $n_-$, are coupled. This results in a positive momentum kick for the COM motion, $e^{i \sqrt{2}k_{\mathrm{soc}} x_{+}}$, if the atoms are in $\ket{\downarrow\downarrow}$, and a negative one, $e^{-i \sqrt{2} k_{\mathrm{soc}} x_{+}}$, if the atoms are in $\ket{\uparrow\uparrow}$. On the other hand, if the particles have different spins, states with the same COM quantum numbers, $n_+$, and different relative-motion quantum number, $n_-$, are coupled, and SOC leads to a relative motion kick with $e^{i \sqrt{2} k_{\mathrm{soc}} x_{-}}$ for $\left(\ket{\mathrm{S}}+\ket{\mathrm{A}}\right)/\sqrt{2}=\ket{\downarrow\uparrow}$ and $e^{- i \sqrt{2} k_{\mathrm{soc}} x_{-}}$ for $\left(\ket{\mathrm{S}}-\ket{\mathrm{A}}\right)/\sqrt{2}=\ket{\uparrow\downarrow}$. Contrary to this, Raman coupling only connects symmetric pseudo-spin states of the same COM and the same relative motional state. For finite interactions these interpretations of the couplings remain, however the interaction term $H_{\mathrm{int}}$ increases the energy of the $\ket{\downarrow\downarrow}$ and $\ket{\uparrow\uparrow}$ states for finite $g$ and of the $\ket{\mathrm{S}}$ state for finite $g_{\uparrow\downarrow}$ (see Figs.~\ref{fig:diag_energylevel}(b) and (c)). These increases can be calculated exactly and are equal to $2\nu\hbar\omega$, where $\nu$ is given by solving
\begin{equation} \label{eq:nu}
    \frac{-\sqrt{2}\hbar\omega a_{\mathrm{ho}}}{g_{ij}}
    =
    \frac{\Gamma(-\nu))}{2\Gamma(-\nu+1/2)}
\end{equation}
for $i,j=\downarrow,\uparrow$ with harmonic oscillator length $a_{\mathrm{ho}}=\sqrt{\hbar/m\omega}$ \cite{Busch1998}. As $2\nu$ is always smaller than 1, the interaction energy shifts do not lead to level crossings or new degeneracies.

As mentioned, the interactions give rise to couplings between different basis states and energy shifts. The couplings cause avoided crossings as shown in Sec.~\ref{sec:results} and Ref.~\cite{Guan2014}. The energy shifts compete with the SOC and generate different effects depending on whether their origin is due to $g$ or $g_{\uparrow\downarrow}$ (see Fig.~\ref{fig:diag_energylevel}).
For non-zero $g_{\uparrow\downarrow}$, the energy gaps for transitions from $\ket{\mathrm{S}}$ to $\ket{\mathrm{A}}$ shrink, while they grow for transitions from $\ket{\mathrm{A}}$ to $\ket{\mathrm{S}}$ (see Fig.~\ref{fig:diag_energylevel}(b)). On the other hand, for non-zero $g$, the energies of states with $\ket{\downarrow\downarrow}$ and $\ket{\uparrow\uparrow}$ all rise by the same amount, which means that the energy gaps between the coupled states stay fixed at $\hbar\omega$ (see Fig.~\ref{fig:diag_energylevel}(c)). The exact strengths of the different contact interactions therefore have well-defined and, in principal different effects on the energy-level structure of the SO-coupled system. We will explore the consequences of these below. 

\section{Results}\label{sec:results}

In the following we will discuss the ground state and energy spectrum of the system for the three cases of no interactions, finite anti-aligned interactions, and finite aligned interactions. The respective ground states will be interpreted by looking at the populations of the different pseudo-spin states, their momentum and density correlations, and their entanglement properties. As an aid to clarity and generality we will use scaled parameters for all plots by giving all energies in units of $\hbar\omega$, all momenta in units of $\hbar/a_{\mathrm{ho}}$, and all interaction strengths in units of $\hbar\omega a_{\mathrm{ho}}$. However, throughout the text we will work with unscaled variables. 

\subsection{Zero interactions $(g=g_{\uparrow\downarrow}=0)$}
For weakly interacting bosons in the mean-field regime, the presence of SOC leads to the possibility of having three different ground state phases \cite{Li2012,Zhai2015}: the stripe phase, the magnetised phase, and the single minimum phase. The stripe phase is named after the fact that an interference pattern appears due to superposition of positive and negative momenta \cite{Li2012}. In contrast, in the magnetised phase the gas either fully adopts positive or negative momentum and in the single minimum phase the spectrum only possess a single minimum. For non-interacting bosons only the stripe phase and the single minimum phase exist. In free space the critical point between these two phases is given by $\hbar\Omega_{\mathrm{c}}=2\hbar^2k_{\mathrm{soc}}^2/m$ and it is known numerically that for harmonically trapped systems the value of the critical point is lower \cite{Zhu2016}.

\begin{figure}[tb]
    \centering
    \includegraphics[width=0.8\textwidth]{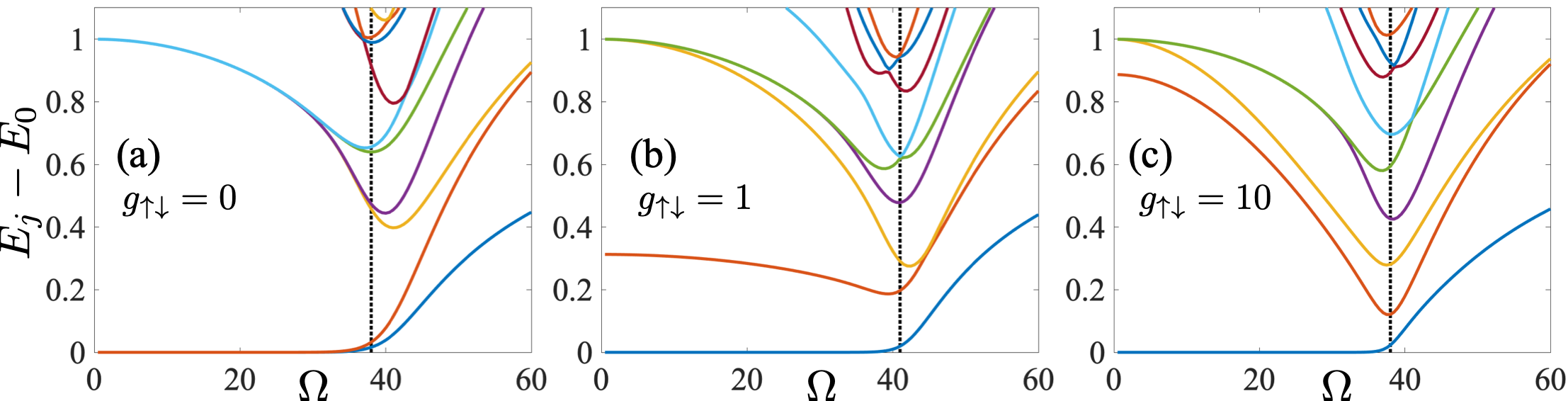}
    \caption{Energy differences $E_j-E_0$ between the $j$-th excited states ($j=1,2,...$) and the ground state, for fixed $k_{\mathrm{soc}}a_{\mathrm{ho}}=5$ and $g/\hbar\omega a_{\mathrm{ho}}=0$. (a) No interactions, $g_{\uparrow\downarrow}/\hbar\omega a_{\mathrm{ho}}=0$; (b) weak interactions, $g_{\uparrow\downarrow}/\hbar\omega a_{\mathrm{ho}}=1$; (c) strong interactions, $g_{\uparrow\downarrow}/\hbar\omega a_{\mathrm{ho}}=10$. The dotted lines indicate the point at which the degeneracies are lifted: (a) $\Omega_{\mathrm{lift}}/\omega\simeq38$ for $g_{\uparrow\downarrow}/\hbar\omega a_{\mathrm{ho}}=0$, (b) $\Omega_{\mathrm{lift}}/\omega\simeq41$ for $g_{\uparrow\downarrow}/\hbar\omega a_{\mathrm{ho}}=1$, and (c) $\Omega_{\mathrm{lift}}/\omega\simeq38$ for $g_{\uparrow\downarrow}/\hbar\omega a_{\mathrm{ho}}=10$. It is defined as the point where $E_0$ and $E_1$ start to deviate, $(E_1-E_0)/\hbar\omega\gtrsim10^{-2}$.
    }
    \label{fig:energyspectrum}
\end{figure}

It is worth noting that in the non-interacting limit the above Hamiltonian is analogous to that of certain collective spin models, namely the Dicke model \cite{Dicke1954} and the Lipkin-Meshkov-Glick (LMG) model \cite{Lipkin1965}. The explicit connection between SO-coupled systems and the Dicke model was established in Refs.~\cite{Huang2015,Hamner2014}, and there is a direct relation between the Dicke and LMG models shown in, for example, Ref.~\cite{Bakemeier2012}. In the thermodynamic limit the ground state of the LMG model is known to exhibit a second order phase transition at a critical value of an applied field~\cite{LMG1,LMG2}, which corresponds to the transition between the stripe phase and the single minimum phase for SO-coupled systems when changing the Raman coupling strength. This transition is signalled by a divergence in the second derivative of the ground state energy and can also be seen in the lifting of existing degeneracies \cite{Campbell2016a}. This duality between the many-body spin system and the SO-coupled atoms emerges due to the nature of the phase transition in the LMG case. There the infinite range of the spin-spin interactions permits expression of the many-body system in terms of collective spin operators with the total angular momentum being conserved. In this picture the LMG model is represented by a single large $N$-dimensional spin, and in thermodynamic limit one phase is effectively given by a double-well configuration possessing a ground state energy degeneracy, while in the other a gaped spectrum exists. This is analogous to the form of the dispersion relation for the stripe phase and the single minimum phase. Naturally, a two-particle system is far from the thermodynamic limit and therefore cannot be expected to exhibit all characteristics of the transition, e.g. the discontinuity in the second derivative of the energy \cite{Hamner2014}. Nevertheless, it is interesting to note that at a value of $\Omega\simeq\Omega_{\mathrm{lift}}$ the degeneracies in the two-particle spectrum are still lifted (see Fig.~\ref{fig:energyspectrum} and cf.~Fig.~1 of \cite{Campbell2016a}). 
 
One can also see from Fig.~\ref{fig:diag_energylevel}(a) that in the absence of Raman coupling the ground state of the two atom system is three-fold degenerate: $\ket{\downarrow\downarrow}$ with COM motion $e^{ik_{\mathrm{soc}}x_+}$ in the positive direction, $\ket{\uparrow\uparrow}$ with COM motion $e^{-ik_{\mathrm{soc}}x_+}$ in the negative direction, and the symmetric combination of $\ket{\downarrow\uparrow}$ and $\ket{\uparrow\downarrow}$ with non-zero relative motion. Finite strengths of the Raman coupling then mix these states, and in the strong Raman coupling limit, $\Omega\gg \Omega_{\mathrm{lift}}$, the ground state becomes an equally-weighted superposition of all pseudo-spin states, $\psi\sim\left(\ket{\downarrow\downarrow}-\ket{\downarrow\uparrow}-\ket{\uparrow\downarrow}+\ket{\uparrow\uparrow}\right)/2$. 

\subsection{Anti-aligned interactions $(g_{\uparrow\downarrow}\!>\!0,~g\!=\!0)$}
Let us next investigate the competition between SOC and repulsive contact interactions between anti-aligned spins, $g_{\uparrow\downarrow}>0$. The level scheme is shown in Fig.~\ref{fig:diag_energylevel}(b) and one can immediately see that the finite interaction lifts a number of degeneracies, as the energy of the basis states of the relative motion is increased.  The spectra resulting from these nontrivial energy shifts for two different interaction strengths are shown in Figs.~\ref{fig:energyspectrum}(b) and (c).

Let us consider $\Omega\ll\Omega_{\mathrm{lift}}$ first. While without interactions the lowest energy states are composed from three degenerate states, for $g_{\uparrow\downarrow}>0$ the energies of $\ket{\downarrow\uparrow}$ and $\ket{\uparrow\downarrow}$ rise, and the lowest energy states are thus the symmetric spin states with positive momentum of COM motion, $\ket{\downarrow\downarrow}e^{ik_{\mathrm{soc}}x_+}$, and negative one, $\ket{\uparrow\uparrow}e^{-ik_{\mathrm{soc}}x_+}$. The first excited state is given by the symmetric superposition of the anti-aligned states, which each have non-zero relative motion, $\psi\sim\ket{\downarrow\uparrow} e^{ik_{\mathrm{soc}}x_-}+\ket{\uparrow\downarrow} e^{-ik_{\mathrm{soc}}x_-}$, and it has an energy shift of $2\nu\hbar\omega$, $\nu$ of which is given by solving Eq.~\eqref{eq:nu} and depends on the interaction strength $g_{\uparrow\downarrow}$. In the strong coupling regime ($\Omega\gg\Omega_{\mathrm{lift}}$), the effect of the interactions becomes negligible, and the ground state approaches that of no interactions. As the states for weak and strong Raman coupling regimes must be adiabatically connected, the crossover from states that are dominated by the interaction to those that are not leads to the appearance in the spectrum of higher-lying avoided crossings \cite{Guan2014}. However, the position of the critical point is only slightly modified by the interactions, due to the fact that the resulting energy shift can at most be $\Delta E\!=\!\hbar\omega$, which is small compared to the energy scales given by the recoil energy and the Raman coupling. 

\begin{figure}[tb]
  \includegraphics[width=\textwidth]{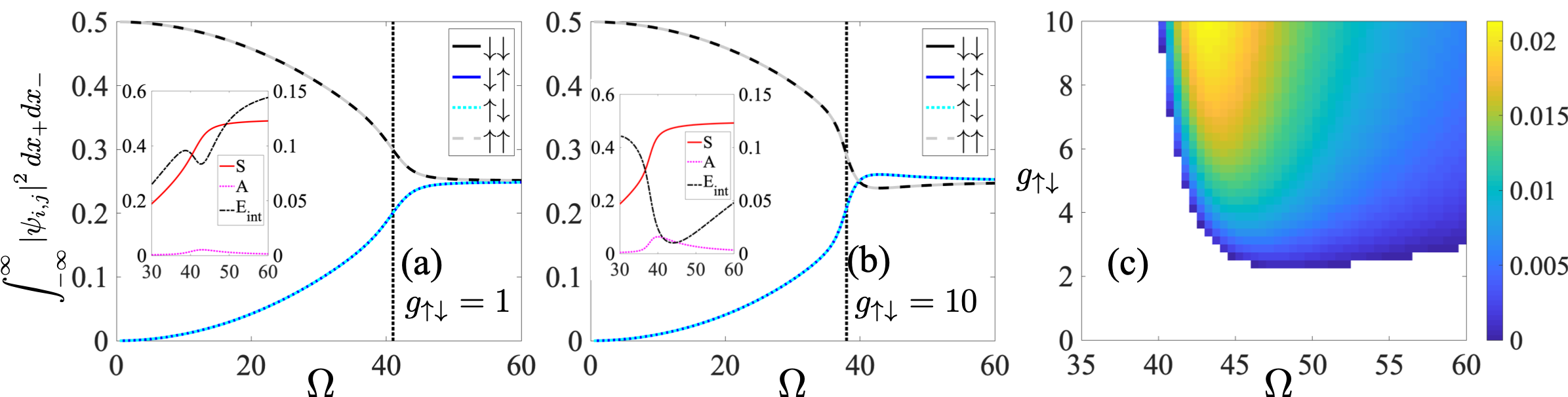}
  \caption{Population of each pseudo-spin state of the ground state with $k_{\mathrm{soc}}a_{\mathrm{ho}}=5$ fixed for (a) weak interactions, $g_{\uparrow\downarrow}/\hbar\omega a_{\mathrm{ho}}=1$ and (b) strong interactions, $g_{\uparrow\downarrow}/\hbar\omega a_{\mathrm{ho}}=10$. The black dotted lines indicate $\Omega_{\mathrm{lift}}/\omega\simeq41$ in (a) and $\Omega_{\mathrm{lift}}/\omega\simeq38$ in (b). The insets show the population of the states $\ket{\mathrm{S}}$ (red line) and $\ket{\mathrm{A}}$ (pink dotted line) using the left axis and the interaction energy (black dashed line) using the right axis. (c) Population difference $\int dx_1 \int dx_2 (|\psi_{\downarrow\uparrow}|^2+|\psi_{\uparrow\downarrow}|^2-|\psi_{\downarrow\downarrow}|^2-|\psi_{\uparrow\uparrow}|^2)$ of the ground state as a function of $\Omega$ and $g_{\uparrow\downarrow}$. Only positive values, where the anti-aligned states dominate, are shown.}
  \label{fig:population}
\end{figure}

The contributions to the ground state of each pseudo-spin state are shown in Fig.~\ref{fig:population}. For weak interactions ($g_{\uparrow\downarrow}/\hbar\omega a_{\mathrm{ho}}=1$, see panel~(a)) and for weak Raman coupling ($\Omega/\omega\lesssim 10$) the ground state is given approximately by $\psi\sim(\ket{\downarrow\downarrow}+\ket{\uparrow\uparrow})/\sqrt{2}$, and is reminiscent of that of the stripe phase in BECs with SOC as each spin component possesses a finite COM momentum \cite{Li2012}. For increasing $\Omega$, a greater proportion of anti-aligned states is mixed into the ground state, however the symmetric states $\{\ket{\downarrow\downarrow}$, $\ket{\mathrm{S}}$, $\ket{\uparrow\uparrow}\}$ always dominate (see inset in panel (a)). This behaviour is similar for systems with strong interactions ($g_{\uparrow\downarrow}/\hbar\omega a_{\mathrm{ho}}=10$, see panel (b)), but one can also see a region in which the contributions from the aligned spins drops faster with increasing $\Omega$, to the point where the contribution from the anti-aligned states exceeds those from aligned states. This occurs around the value of $\Omega=\Omega_{\mathrm{lift}}$  and we show in panel~(c) that this inversion appears already for interaction strengths $g_{\uparrow\downarrow}/\hbar\omega a_{\mathrm{ho}}\!\gtrsim\!2$. The phase after the inversion is not analogous to any phase of BECs with SOC in the weakly interacting mean-field limit and we will discuss it in more detail below. Increasing $\Omega$ even further, the ground state again moves towards that corresponding to that for the single minimum phase.

While one would naively assume that the ground state contains more interaction energy when larger contributions of the $\ket{\downarrow\uparrow}$ and $\ket{\uparrow\downarrow}$ states are present, one can see from the inset in panel (b) that the increase in contributions from the anti-aligned states is due to an increase in the populations of the anti-symmetric state $\ket{\mathrm{A}}$, which does not feel the interaction \cite{Guan2015}. In fact, when the contribution of $\ket{\mathrm{A}}$ rises, the interaction energy can be seen to decrease (insets in panels ~(a,b)). This population imbalance inversion is a few-body effect and is not seen when treating single-particle states or BECs within the mean-field approximation. We therefore refer to the distinct and unique regime after this inversion as the AS ground state phase, as the anti-symmetric states are the dominant contribution.


\begin{figure}[tb]
  \includegraphics[width=\textwidth]{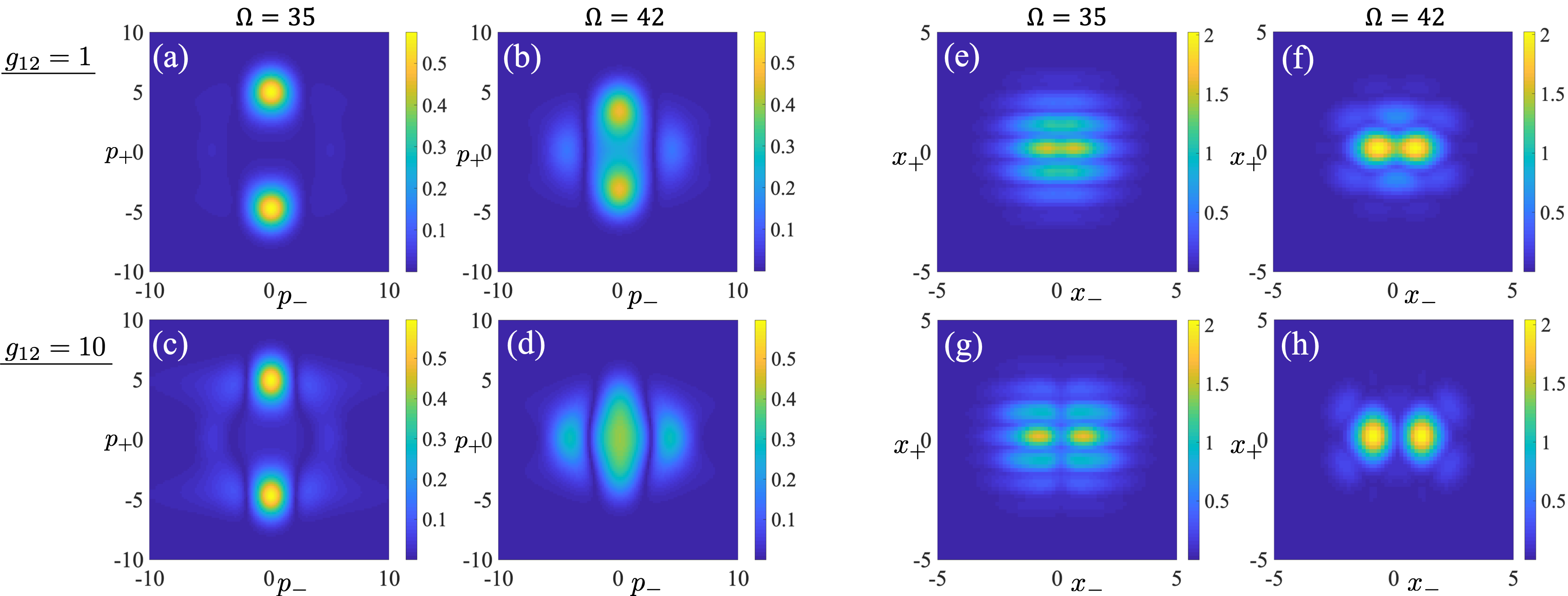}
  \centering
  \caption{
  Overall momentum distributions (left) and overall density distributions (right) for weak interaction $g_{\uparrow\downarrow}/\hbar\omega a_{\mathrm{ho}}=1$ (upper row) and for strong interaction $g_{\uparrow\downarrow}/\hbar\omega a_{\mathrm{ho}}=10$ (lower row) with $k_{\mathrm{soc}}a_{\mathrm{ho}}=5$ fixed. In panels (a,c,e,g), $\Omega/\omega=35$, and in panels (b,d,f,h) $\Omega/\omega=42$, which are below and above the value for which the inversion occurs.
  }
  \label{fig:correlation_momden}
\end{figure}

The crossover from a ground state resembling the stripe phase to the AS ground state has direct consequences on the overall momentum and density distributions (see Fig.~\ref{fig:correlation_momden}). For weak interaction ($g_{\uparrow\downarrow}/\hbar\omega a_{\mathrm{ho}}=1$, see panels~(a,b)), when the AS state does not appear, the ground state momentum distribution is dominated by a finite COM momentum on both sides of $\Omega_\text{lift}$. However, for a strong interaction ($g_{\uparrow\downarrow}/\hbar\omega a_{\mathrm{ho}}=10$, see panels~(c,d)) this distribution changes from being dominated by the COM contribution to having more equal contributions from the COM and relative momenta, which is a sign that considerable amounts of $\ket{\downarrow\uparrow}$ and $\ket{\uparrow\downarrow}$ appear. For the density distribution an interference pattern along the COM coordinate is visible for $\Omega\lesssim\Omega_\text{lift}$ (see panels~(e,g)), which is reminiscent of the stripe phase, but with a bisection due to the contact interaction at $x_-=0$. As $\Omega$ increases, the interference pattern remains for a weak interaction (see panel~(f)), however, for a strong interaction a more pronounced pattern in the direction of the relative coordinate appears while that in the COM direction vanishes (see panel~(h)). These changes in the spatial and momentum distributions clearly indicate the appearance of the AS ground state. 

We next turn our attention to the non-classical correlations inherent in the system. For this we use the concurrence as the entanglement measure of the pseudo-spin degree of freedom, which is constructed from the density matrix after tracing over the position space components. Representing the wave function of the ground state as $\ket{\psi}=\sum_{\chi}\phi_{\chi}(x_1,x_2)\ket{\chi}$ for $\chi\in\{ \downarrow\downarrow, \downarrow\uparrow, \uparrow\downarrow, \uparrow\uparrow \}$, the density matrix can be written as $\rho = \vert \psi \rangle \langle \psi \vert$, and the reduced density matrix is real and given by 
\begin{equation} \label{eq:rho_spin}
    \rho_{\mathrm{spin}} =
    \begin{pmatrix}
        \theta &
        \beta &
        \beta &
        \gamma \\
        \beta & 
        \epsilon & 
        \mu & 
        \beta \\
        \beta  &
        \mu  & 
        \epsilon & 
        \beta \\
        \gamma &
         \beta &
         \beta & 
        \theta
    \end{pmatrix}\;.
\end{equation}
The inner products of the two aligned components are given by $\theta=\langle\phi_{\downarrow\downarrow} \vert \phi_{\downarrow\downarrow} \rangle =\langle\phi_{\uparrow\uparrow} \vert \phi_{\uparrow\uparrow} \rangle$ and $\gamma=\langle\phi_{\downarrow\downarrow} \vert \phi_{\uparrow\uparrow} \rangle =\langle\phi_{\uparrow\uparrow} \vert \phi_{\downarrow\downarrow} \rangle$, while the inner products of two anti-aligned components are given by $\epsilon=\langle\phi_{\downarrow\uparrow} \vert \phi_{\downarrow\uparrow} \rangle =\langle\phi_{\uparrow\downarrow} \vert \phi_{\uparrow\downarrow} \rangle$ and $\mu=\langle\phi_{\downarrow\uparrow} \vert \phi_{\uparrow\downarrow} \rangle =\langle\phi_{\uparrow\downarrow} \vert \phi_{\downarrow\uparrow} \rangle$. The diagonal elements of the density matrix therefore describe the population of each pseudo-spin state. Finally, the inner products of one aligned and one anti-aligned component are all equivalent and given by $\beta=\langle\phi_{\downarrow\downarrow} \vert \phi_{\downarrow\uparrow} \rangle =\langle\phi_{\downarrow\downarrow} \vert \phi_{\uparrow\downarrow} \rangle=\langle\phi_{\uparrow\uparrow} \vert \phi_{\uparrow\downarrow}\rangle=\langle\phi_{\uparrow\uparrow} \vert \phi_{\downarrow\uparrow} \rangle$ and the same for their Hermitian conjugates.

\begin{figure}[tb]
    \includegraphics[width=0.9\textwidth]{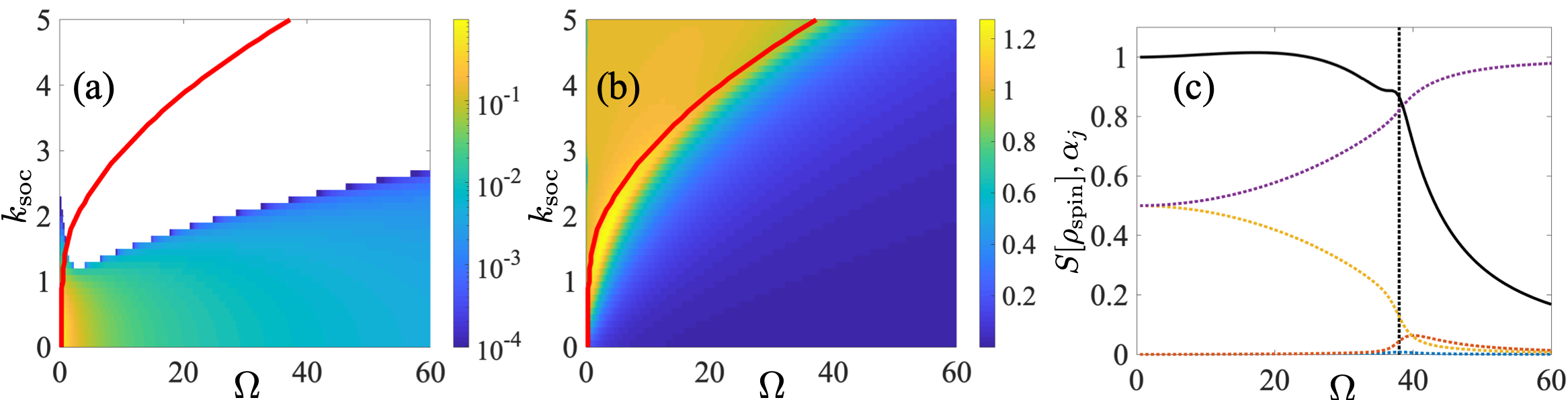}
    \centering
    \caption{(a) Concurrence and (b) vNE of the ground state for a range of $k_{\mathrm{soc}}a_{\mathrm{ho}}$  with $g_{\uparrow\downarrow}/\hbar\omega a_{\mathrm{ho}}=10$ fixed. The red lines indicate the value of $\Omega_{\mathrm{lift}}$ for each $k_{\mathrm{soc}}$ and in (a) the white area corresponds to $C_{\mathrm{spin}}=0$. (c) vNE as a function of $\Omega$ for $k_{\mathrm{soc}}a_{\mathrm{ho}}=5$ fixed (black line). The coloured, dotted lines represent the values of the eigenvalues $\alpha_j$ of the reduced density matrix $\rho_{\mathrm{spin}}$ for $j=1,2,3,4$. The black dotted line indicates $\Omega_{\mathrm{lift}}/\omega\simeq38$.
    \label{fig:concurrence}}
\end{figure}

The concurrence is then given as
\begin{equation}
    C_{\mathrm{spin}} = \max \{0,\sqrt{\lambda_1}-\sqrt{\lambda_2}-\sqrt{\lambda_3}-\sqrt{\lambda_4}\}\;,
\end{equation}
where the $\lambda_j$ are the eigenvalues of $\rho_{\mathrm{spin}}(\sigma_y\otimes\sigma_y)\rho_{\mathrm{spin}}^{*}(\sigma_y\otimes\sigma_y)$ in descending order for $j=1,2,3,4$ \cite{Wootters1998,Wootters2001}. For separable states the concurrence vanishes, and for maximally entangled states it is equal to one. In Fig.~\ref{fig:concurrence}(a) we show the behaviour of the concurrence as a function of Raman coupling and SOC strengths for a fixed contact interaction, $g_{\uparrow\downarrow}/\hbar\omega a_{\mathrm{ho}}=10$. In the limit of $k_{\mathrm{soc}},\Omega\to 0$, the ground state is given by $\left(\ket{\downarrow\downarrow}+\ket{\uparrow\uparrow}\right)/\sqrt{2}$, and hence $C_{\mathrm{spin}} \! \to \! 1$, i.e.~the maximally entangled Bell state~\cite{SchmiedmayerPRA}. It is a key point that the diagonal terms $\theta$ and the off-diagonal terms $\gamma$ of the reduced density matrix \eqref{eq:rho_spin} go to 1/2 and the remaining inner products disappear in this limit. The off-diagonal terms $\gamma$ show the correlation between $\ket{\downarrow\downarrow}$ and $\ket{\uparrow\uparrow}$. However for finite $k_{\mathrm{soc}}$ the entanglement decays for all values of $\Omega$, which is despite the fact that for small $\Omega$ the ground state is equally weighted between $\ket{\downarrow\downarrow}$ and $ \ket{\uparrow\uparrow}$, and therefore one would expect a maximally entangled state. The entanglement decays because the momentum provided by SOC suppresses the off-diagonal terms $\gamma$ and kill the correlation. The SOC term therefore acts as an effective dephasing channel on the pseudo-spin states causing the coherences to decay rapidly while the diagonal terms, which account for the populations, remain unaffected. A final remarkable feature of the entanglement is its sudden vanishing, a phenomenon sometimes referred to as entanglement sudden death~\cite{ESD}. With the red line indicating the values at which the degeneracies are lifted (see Fig.~\ref{fig:concurrence}(a)) one can immediately see that the pseudo-spin entanglement is not a useful indicator of the population inversion. However the effective dephasing caused by the SOC also implies that strong correlations are established between the pseudo-spin space and  real-space.

As the overall state is pure, we can analyse this entanglement using the von Neumann entropy (vNE) of the reduced density matrix $\rho_{\mathrm{spin}}$, defined as
\begin{equation}
\begin{split}
    S[\rho_{\mathrm{spin}}]
    &= -\mathrm{Tr}\left\{\rho_{\mathrm{spin}}\log_2 \rho_{\mathrm{spin}}\right\} \\
    &= -\sum_{j=1}^4 \alpha_j \log_2 \alpha_j,
\end{split}
\end{equation}
where $\alpha_j$ is the $j$-th eigenvalue of the reduced density matrix, $\rho_\text{spin}\phi_j=\alpha_j \phi_j$. 
The behaviour of the vNE can be linked to the pseudo-spin populations through the eigenvalues of $\rho_{\mathrm{spin}}$ (see Fig.~\ref{fig:concurrence}(c)), whereby we can explicitly calculate two eigenvalues as $\alpha_1=\theta-\gamma=\langle \phi_{\downarrow \downarrow} \vert \phi_{\downarrow \downarrow}\rangle-\langle \phi_{\downarrow \downarrow} \vert \phi_{\uparrow \uparrow} \rangle$ (yellow dotted) and $\alpha_2=\epsilon-\mu=\langle \phi_{\downarrow \uparrow} \vert \phi_{\downarrow \uparrow} \rangle-\langle \phi_{\downarrow \uparrow} \vert \phi_{\uparrow \downarrow} \rangle$ (orange dotted). Also, noticing that $\alpha_3\approx0$ (blue dotted) allows us to approximate the last eigenvalue as $\alpha_4\approx \theta + \gamma +\epsilon + \mu= \langle \phi_{\downarrow \downarrow}\vert \phi_{\downarrow \downarrow} \rangle +\langle \phi_{\downarrow \downarrow}\vert \phi_{\uparrow \uparrow}\rangle +\langle \phi_{\downarrow \uparrow}\vert\phi_{ \downarrow \uparrow}\rangle+\langle \phi_{\downarrow \uparrow}\vert \phi_{\uparrow \downarrow }\rangle$ (purple dotted). In the weak Raman coupling limit, $\Omega\ll\Omega_{\mathrm{lift}}$, the vNE remains close to 1, which is half its maximal value (see Fig.~\ref{fig:concurrence}(b) and (c)). That is to be expected as the $\ket{\downarrow\downarrow}$ state and $\ket{\uparrow\uparrow}$ state, which dominate the system, are connected to COM motion in the positive and negative directions, respectively, while the other two pseudo-spin states, $\ket{\downarrow\uparrow}$ and $\ket{\uparrow\downarrow}$, have lower populations and do not play major roles in the correlations. For Raman coupling strengths closer to $\Omega_{\mathrm{lift}}$ one can see a kink appear in the behaviour of the vNE, which can be directly linked to the increase in population of the $\ket{\mathrm{A}}$ state. This is also signalled by the eigenvalue $\alpha_2$ becoming non-zero, 
which implies the establishment of entanglement between the $\ket{\downarrow\uparrow}$ and $\ket{\uparrow\downarrow}$ states. For $\Omega\gtrsim\Omega_{\mathrm{lift}}$ the vNE decreases dramatically (the red line in Fig.~\ref{fig:concurrence}(b) indicates $\Omega=\Omega_{\mathrm{lift}}$) as the system possesses an almost equal contribution of all the pseudo spin states in the AS phase. Finally, when the wavelength of the SOC is on the order of the width of the ground state, $\sqrt{2}k_{\mathrm{soc}}\sim 1/a_{\mathrm{ho}}$, the vNE becomes maximal due to an enhanced coupling between the spin states as a result of the finite system size.  

\subsection{Aligned interaction, $g>0,~g_{\uparrow\downarrow}=0$}
\begin{figure}[tb]
\includegraphics[width=\textwidth]{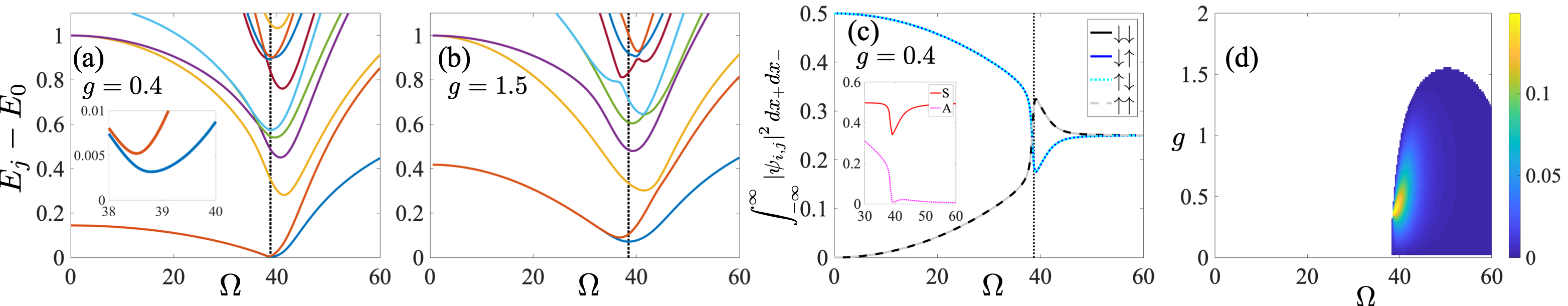}
\caption{(a,b) Energy differences $E_j-E_0$ between the $j$th excited states ($j=1,2,...$) and the ground state with $k_{\mathrm{soc}}a_{\mathrm{ho}}=5$ fixed for $g/\hbar\omega a_{\mathrm{ho}}=0.4$ and $g/\hbar\omega a_{\mathrm{ho}}=1.5$, respectively. The dotted lines represent the value of $\Omega_{\mathrm{lift}}/\omega$, which is defined as the point where $E_1$ and $E_2$ starts to deviate, $(E_2-E_1)/\hbar\omega\gtrsim10^{-2}$, and which is very close to the point where the avoided crossing between the ground state and the first excited state appears, $\Omega_{\mathrm{lift}}/\omega\simeq39$. (c) Population of each pseudo-spin state of the ground state with $g/\hbar\omega a_{\mathrm{soc}}=0.4$ fixed. (d) Population difference $\int dx_1 \int dx_2 (|\psi_{\downarrow\uparrow}|^2+|\psi_{\uparrow\downarrow}|^2-|\psi_{\downarrow\downarrow}|^2-|\psi_{\uparrow\uparrow}|^2)$ of the ground state as a function of $\Omega$ and $g$. Only positive values, where the aligned states dominate, are shown.}
\label{fig:population_g}
\end{figure}

Finally we investigate the case where only the aligned interactions are finite, while the anti-aligned interactions are switched off (see Fig.~\ref{fig:diag_energylevel}(c)). In this situation, the basis states $\ket{\uparrow\uparrow}$ and $\ket{\downarrow\downarrow}$ are shifted in energy, while the symmetric $\ket{\mathrm{S}}$ and anti-symmetric $\ket{\mathrm{A}}$ states are unaffected, resulting in a different ground state compared to that seen for anti-aligned interactions. The energy spectra are shown in Figs.~\ref{fig:population_g}(a) and (b), and one can immediately confirm that for all values of $\Omega$ the ground state is non-degenerate (see inset in panel (a)). In fact, for small $\Omega$ it is given by $\psi\sim \ket{\downarrow\uparrow} e^{ik_{\mathrm{soc}}x_-}+\ket{\uparrow\downarrow} e^{-ik_{\mathrm{soc}}x_-}$ and unsurprisingly composed mostly of anti-aligned pseudo-spin states. The first excited states are composed of two degenerate states with COM momentum, $\psi\sim\ket{\downarrow\downarrow} e^{ik_{\mathrm{soc}}x_+}$ and $\psi\sim\ket{\uparrow\uparrow} e^{-ik_{\mathrm{soc}}x_+}$, 
and one can see that this degeneracy is resolved at a critical value similar to the one for the finite anti-aligned interactions. Interestingly, for weak aligned interactions ($g/\hbar\omega a_{\mathrm{ho}}=0.4$, see the inset in Fig.~\ref{fig:population_g}(a)) an avoided crossing between the ground state and the first excited states exists when 
$\Omega/\omega\simeq39\simeq \Omega_{\mathrm{lift}}/\omega$. This avoided crossing is the result of the interaction and the Raman coupling  trying to shift the energy in opposite directions \cite{Guan2014}.  
For stronger interactions ($g/\hbar\omega a_{\mathrm{soc}}=1.5$, see Fig.~\ref{fig:population_g}(b)), the energy gap at the avoided crossing is increased as the contact interactions push the excited states to higher energies.
The presence of the avoided crossing also impacts the population imbalance after the inversion. Compared to the case of the anti-aligned interactions, the inversion from the anti-aligned basis states being the dominant contributions to the ground state to the aligned ones now occurs more sharply (see Fig.~\ref{fig:population_g}(c)). Although the inversion does not occur without the interaction, it is worth noting that it appears even for an infinitesimal weak interaction and decays  for increasing $g$ (see Fig.~\ref{fig:population_g}(d)). This is a manifestation of a non-zero $g$ shifting energy levels differently from non-zero $g_{\uparrow\downarrow}$. For the case of finite anti-aligned interaction, populating the $\ket{\mathrm{A}}$ state reduces the interaction energy (see Fig.\ref{fig:diag_energylevel}(b)), whereas for non-zero $g$ increasing $\Omega$ leads to larger contributions of $\ket{\downarrow\downarrow}$ and $\ket{\uparrow\uparrow}$ to the ground state (see Fig.\ref{fig:diag_energylevel}(c)). However, these states both contribute to the interaction energy, and therefore for large $g$ they are no longer favourable and the inversion disappears.

\begin{figure}[tb]
\includegraphics[width=\textwidth]{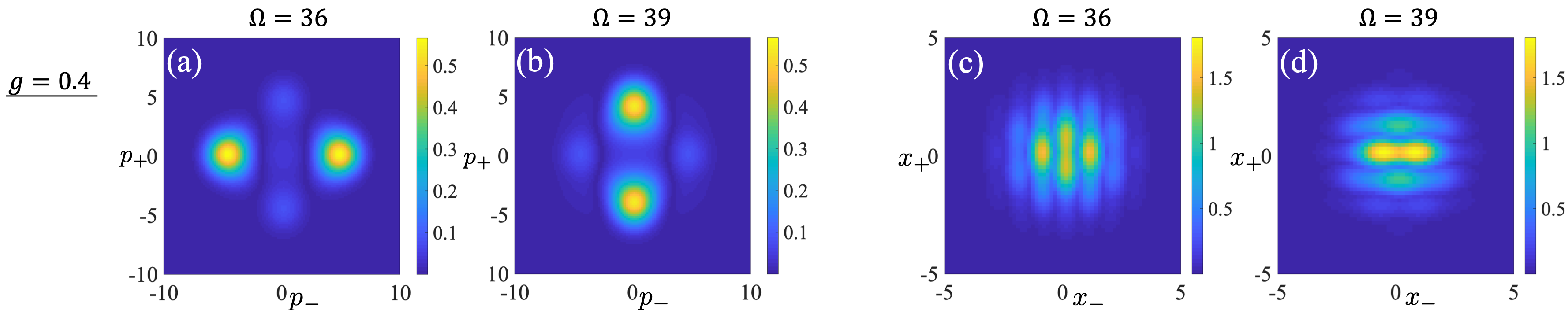}
\caption{
Overall momentum distributions (a,b) and density distributions (c,d) for $g/\hbar\omega a_{\mathrm{ho}}=0.4$ with $k_{\mathrm{soc}}a_{\mathrm{ho}}=5$ fixed. In panels~(a,c), $\Omega/\omega=36$, and in panels~(c,d) $\Omega/\omega=39$, which are just below and above the value for which the inversion occurs, respectively.
}
\label{fig:correlation_denmom_g}
\end{figure}

The effect of the inversion on the overall momentum distribution is shown in Fig.~\ref{fig:correlation_denmom_g}(a,b). For $\Omega\ll\Omega_{\mathrm{lift}}$, the largest contribution to the ground state is the anti-aligned states $\ket{\downarrow\uparrow}$ and $\ket{\uparrow\downarrow}$ which possess net relative momentum from the SOC. When $\Omega\simeq\Omega_{\mathrm{lift}}$, the population inversion occurs and the ground state possesses net COM momentum in the majority aligned states $\ket{\downarrow\downarrow}$ and $\ket{\uparrow\uparrow}$. This corresponds directly to the overall density distribution exhibiting a reorientation of the interference fringes from the relative to the COM direction in the crossover region (see Fig.~\ref{fig:correlation_denmom_g}(c,d)).

\begin{figure}[t]
\includegraphics[width=0.8\textwidth]{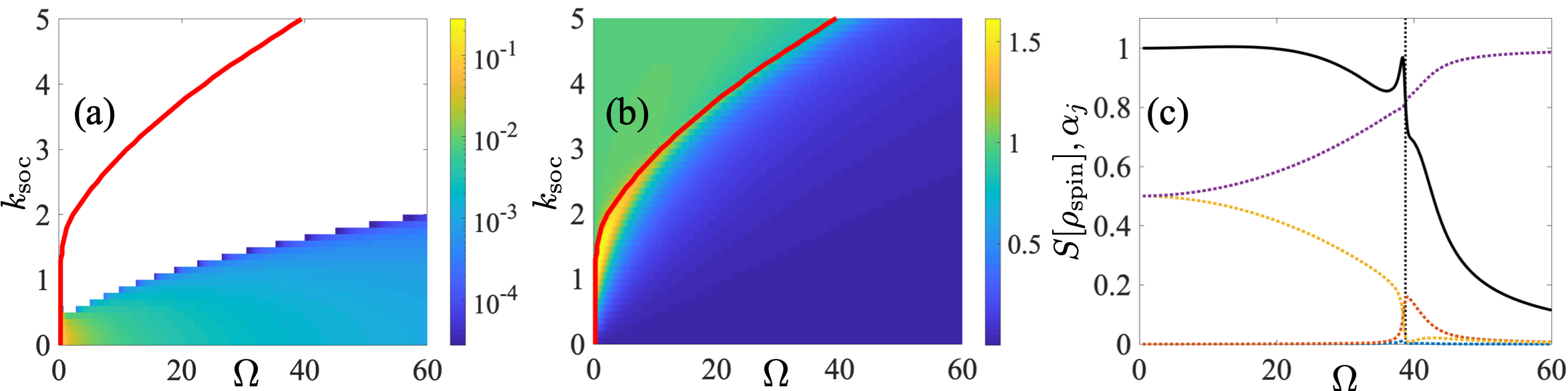}
\centering
\caption{
(a) Concurrence and (b) vNE of the ground state for a range of $k_{\mathrm{soc}}$ with $g/\hbar\omega a_{\mathrm{ho}}=0.4$ and $k_{\mathrm{soc}}a_{\mathrm{ho}}=5$ fixed. The white domain shows $C_{\mathrm{spin}}=0$. The red lines represent $\Omega_{\mathrm{lift}}/\omega$. (c) vNE with $k_{\mathrm{soc}}a_{\mathrm{ho}}=5$ fixed. The maximal value is given by $\mathrm{max}[S[\rho_{\mathrm{spin}}]]=2$. The coloured, dotted lines represent the eigenvalues $\alpha_j$ of the reduced density matrix $\rho_{\mathrm{spin}}$ for $j=1,2,3,4$. The black dotted line indicates $\Omega_{\mathrm{lift}}/\omega\simeq39$.}
\label{fig:concurrence_g}
\end{figure}

The concurrence $C_{\mathrm{spin}}$ and the vNE $S[\rho_{\mathrm{spin}}]$ for $g/\hbar\omega a_{\mathrm{soc}}=0.4$, where the population inversion is large, can be seen in Fig.~\ref{fig:concurrence_g}(a,b). Simlar to the case of finite anti-aligned interactions, the concurrence does not signal the lifting of the energy degeneracy, but the vNE significantly decreases at $\Omega\simeq\Omega_{\mathrm{lift}}$. Furthermore, a sharp spike is observed in the vNE whenever $\Omega/\omega\simeq\Omega_{\mathrm{lift}}$ (see Fig.~\ref{fig:concurrence_g}(c) for $k_{\mathrm{soc}}a_{\mathrm{ho}}=5$.)
This is the result of the avoided crossing present in Fig.~\ref{fig:population_g}(a) which strongly entangles $\ket{ \uparrow \uparrow }$ and $\ket{ \downarrow \downarrow }$ spin states at the resonance point as their respective populations suddenly increase, while the symmetric $\ket{\mathrm{S}}$ and anti-symmetric $\ket{\mathrm{A}}$ state populations are suppressed. Indeed, this is the opposite effect described for anti-aligned interactions, and shows how the critical point can enhance correlations between interacting spin-components due to the competition between SOC and contact interactions.

\section{\label{sec:conclusion}Conclusion}
In this work we have considered the effects of repulsive interactions on two SO-coupled particles in a harmonic trap. The finite interactions raise the energy of certain states and therefore lift some of the degeneracies in the free spectrum. We have shown that in the presence of interactions between the anti-aligned spins a ground state unique to strongly correlated systems appears, which is defined by a possessing a finite component of the anti-symmetric spin basis state. The appearance of this state can be directly observed through changes in the density and momentum distributions, but also in the entanglement between the spin and the real space. In the case where the system possess only interactions between aligned spin, a similar inversion in the dominant contribution around a similar value for the Raman coupling strength can be observed, however this time towards the contributions of the aligned spins.

This work helps to bridge the gap between single and many-body systems, and our analysis highlights the role that symmetries play in the energetic and entanglement characteristics of SOC systems. Furthermore, our framework can be used to study how SOC affects non-equilibrium dynamics and will, for example, allow for the efficient simulation of the dynamical generation of entanglement. Indeed, this work has revealed that there is a clear trade-off between the loss of entanglement between pseudo-spins and the generation of strong entanglement between pseudo-spin and real space. This fact indicates that SOC can be used as a tunable knob for creating or distributing entanglement in certain degrees of freedom, and can function as a control parameter in dynamical processes. In addition, we have shown that the SOC term behaves as a pure dephasing channel, allowing for the possibility that SOC systems present viable platforms for the study of certain controlled open quantum systems.

While the population inversion is clearly related to the presence of strong correlations in the system, it is an interesting question to consider systems made from more than two atoms. Since in the case of interactions between anti-aligned states the existence of the anti-symmetric state is the reason for the inversion, it is clear that the symmetry of the system in terms of particle exchange matters. Thus, it is unclear whether the population inversion remains when the particle number is increased. Further investigation into larger particle numbers will therefore require careful investigations and we expect that it is not straightforward to make general predictions.

Finally, with the recent progress in preparing few-particle systems experimentally with high fidelities \cite{Serwane2011,Wenz2013} and new methods to measure their momentum distribution \cite{Bergschneider2018}), we believe that the observation of our predictions is experimentally realistic.

All the code for this paper is available online at Ref.~\cite{code}.

\section{Acknowledgements}
This work has been supported by the Okinawa Institute of Science and Technology Graduate University and used the computing resources of the Scientific Computing and Data Analysis section. T.F. acknowledges funding support under JSPS KAKENHI-18K13507. S.C. gratefully acknowledges the Science Foundation Ireland Starting Investigator Research Grant ``SpeedDemon'' (No.~18/SIRG/5508) for financial support. S.A.G. acknowledges UK EPSRC Grant number EP/R002061/1. The authors thank Yongping Zhang for stimulating discussions.

\appendix
\section*{Appendix} \label{app:matH}
\begin{figure}[tb]
\includegraphics[width=0.4\textwidth]{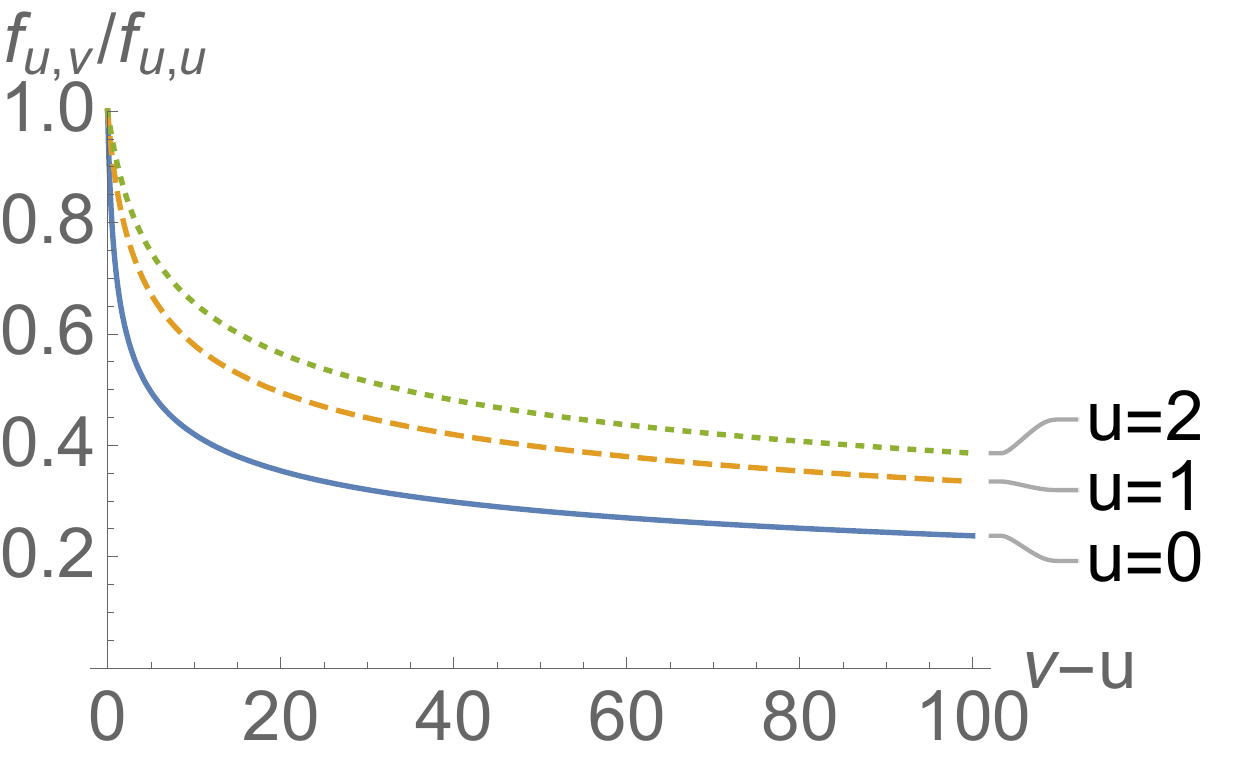}
\centering
\caption{Plot of $f_{u,\nu}/f_{u,u}$, each of which is given by \eqref{eq:fuv}. $f_{u,\nu}$ quantifies strength of coupling caused by contact interactions. The decay of $f_{u,\nu}$ to zero is slow as seen, which requires one to consider sufficient energy levels to converge the system.}
\label{fig:fuv}
\end{figure}

\setcounter{section}{1}
In this appendix we detail the matrix representation of the Hamiltonian for the two interacting bosons in a harmonic trap in the presence of SOC used for the calculations in this work. Starting by ordering the basis states of the harmonic oscillator according to their total energy and value of $n_+$
\begin{equation}
\begin{split}
&   \ket{0,0,\downarrow\downarrow},
    \ket{0,0,\mathrm{S}},
    \ket{0,0,\uparrow\uparrow},
    \ket{0,1,\mathrm{A}},
    \ket{1,0,\downarrow\downarrow},
    \ket{1,0,\mathrm{S}},
    \ket{1,0,\uparrow\uparrow},
    \ket{1,1,\mathrm{A}}
    , \\
&   
    \ket{0,2,\downarrow\downarrow},
    \ket{0,2,\mathrm{S}},
    \ket{0,2,\uparrow\uparrow},
    \ket{0,3,\mathrm{A}}
    ,   
    \ket{2,0,\downarrow\downarrow},
    \ket{2,0,\mathrm{S}},
    \ket{2,0,\uparrow\uparrow},
    \ket{2,1,\mathrm{A}}
    , \\
&   \ket{1,2,\downarrow\downarrow},
    \ket{1,2,\mathrm{S}},
    \ket{1,2,\uparrow\uparrow},
    \ket{1,3,\mathrm{A}},
    \ket{3,0,\downarrow\downarrow},
    \ket{3,0,\mathrm{S}},
    \ket{3,0,\uparrow\uparrow},
    \ket{3,1,\mathrm{A}},  
    \cdots,\nonumber 
\end{split}
\end{equation}
one can produce a matrix representation of the Hamiltonian excluding the interaction term $H_{\mathrm{int}}$ as
\begin{equation}
 H - H_{\mathrm{int}}= 
\begin{pmatrix}
A_{0,0} & B_{1} & C_{2} & 0 & 0 & 0 & 0 & 0 & 0 & \cdots     
\\
B_{1}^{\dagger} & A_{1,0} & 0 & B_{2} & C_{2} & 0 & 0 & 0 & 0 & \cdots
\\
C_{2}^{\dagger} & 0 & A_{0,2} & 0 & B_{1} & 0 & C_{4} & 0 & 0 & \cdots
\\
0 & B_{2}^{\dagger} & 0 & A_{2,0} & 0 & B_{3} & 0 & C_{2} & 0 & \cdots
\\
0 & C_{2}^{\dagger} & B_{1}^{\dagger} & 0 & A_{1,2} & 0 & 0 & B_{2} & 0 & \cdots
\\
0 & 0 & 0 & B_{3}^{\dagger} & 0 & A_{3,0} & 0 & 0 & B_{4} & \cdots
\\
0 & 0 & C_{4}^{\dagger} & 0 & 0 & 0 & A_{0,4} & 0 & 0 & \cdots
\\
0 & 0 & 0 & C_{2}^{\dagger} & B_{2}^{\dagger} & 0 & 0 & A_{2,2} & 0 & \cdots
\\
0 & 0 & 0 & 0 & 0 & B_{4}^{\dagger} & 0 & 0 & A_{4,0} & \cdots
\\
\vdots & \vdots & \vdots & \vdots & \vdots & \vdots & \vdots & \vdots & \vdots & \ddots
\end{pmatrix},
\end{equation}
where each element is a four-by-four matrix such that
\begin{equation}
    A_{n,2u} =
    \begin{pmatrix}
        n+2u+1 & \Upsilon & 0 & 0 \\
        \Upsilon & n+2u+1 & \Upsilon & i\Lambda\sqrt{2u+1} \\
        0 & \Upsilon & n+2u+1 & 0 \\
        0 & -i\Lambda\sqrt{2u+1} & 0 & n+2u+2
    \end{pmatrix}
\end{equation}
\begin{equation}
    B_{n} =
    \begin{pmatrix}
        i\Lambda\sqrt{n} & 0 & 0 & 0 \\
        0 & 0 & 0 & 0 \\
        0 & 0 & -i\Lambda\sqrt{n} & 0 \\
        0 & 0 & 0 & 0
    \end{pmatrix}
\end{equation}
\begin{equation}
    C_{2u} =
    \begin{pmatrix}
        0 & 0 & 0 & 0 \\
        0 & 0 & 0 & 0 \\
        0 & 0 & 0 & 0 \\
        0 & i\Lambda\sqrt{2u} & 0 & 0
    \end{pmatrix}
\end{equation}
for integer $n,u\geq0$. A more compact representation, which also makes the patterns of matrix elements clear, is given by
\begin{equation} \label{eq:Hsp_compact}
    H - H_{\mathrm{int}} = 
    \begin{pmatrix}
        \mathcal{A}_0 & \mathcal{B}_1 & \mathcal{C}_2 & 0 & 0 & 0 & \cdots \\
        \mathcal{B}_1^{\dagger} & \mathcal{A}_1 & \mathcal{B}_2 & \mathcal{C}_2 & 0 & 0 & \cdots \\
        \mathcal{C}_2^{\dagger} & \mathcal{B}_2^{\dagger} & \mathcal{A}_2 & \mathcal{B}_3 & \mathcal{C}_4 & 0 & \cdots \\
        0 & \mathcal{C}_2^{\dagger} & \mathcal{B}_3^{\dagger} & \mathcal{A}_3 & \mathcal{B}_4 & \mathcal{C}_4 & \cdots \\
        0 & 0 & \mathcal{C}_4^{\dagger} & \mathcal{B}_4^{\dagger} & \mathcal{A}_4 & \mathcal{B}_5 & \cdots \\
        0 & 0 & 0 & \mathcal{C}_4^{\dagger} & \mathcal{B}_5^{\dagger} & \mathcal{A}_5 & \cdots \\
        \vdots & \vdots & \vdots & \vdots & \vdots & \vdots & \ddots
    \end{pmatrix},
\end{equation}
where for even and odd indices
\begin{equation}
    \mathcal{A}_{2N} = 
    \begin{pmatrix}
        A_{0,2N} & 0 & \cdots & 0 & 0 \\
        0 & A_{2,2(N-1)} & \cdots & 0 & 0 \\
        \vdots & \vdots & \ddots & \vdots & \vdots \\
        0 & 0 & \cdots & A_{2(N-1),2} & 0 \\
        0 & 0 & \cdots & 0 & A_{2N,0}
    \end{pmatrix},
\end{equation}
\begin{equation}
    \mathcal{A}_{2N+1} = 
    \begin{pmatrix}
        A_{1,2N} & 0 & \cdots & 0 & 0 \\
        0 & A_{3,2(N-1)} & \cdots & 0 & 0 \\
        \vdots & \vdots & \ddots & \vdots & \vdots \\
        0 & 0 & \cdots & A_{2(N-1)+1,2} & 0 \\
        0 & 0 & \cdots & 0 & A_{2N+1,0}
    \end{pmatrix},
\end{equation}
\begin{equation}
      \mathcal{B}_{2N} = 
    \begin{pmatrix}
        0 & B_2 & 0 & \cdots & 0 & 0 \\
        0 & 0& B_{3} & \cdots & 0 & 0 \\
        \vdots & \vdots & \vdots & \ddots & \vdots & \vdots \\
        0 & 0 & 0 & \cdots & B_{2(N-1)} & 0 \\
        0 & 0 & 0 & \cdots & 0 & B_{2N}
    \end{pmatrix},
\end{equation}
\begin{equation}
    \mathcal{B}_{2N+1} = 
    \begin{pmatrix}
        B_{1} & 0 & \cdots & 0 & 0 \\
        0 & B_{3} & \cdots & 0 & 0 \\
        \vdots & \vdots & \ddots & \vdots & \vdots \\
        0 & 0 & \cdots & B_{2(N-1)+1} & 0 \\
        0 & 0 & \cdots & 0 & B_{2N+1}
    \end{pmatrix},
\end{equation}
\begin{equation}
    \mathcal{C}_{2N} = 
    \begin{pmatrix}
        C_{2N} & 0 & \cdots & 0 & 0 & 0 \\
        0 & C_{2(N-1)} & \cdots & 0 & 0 & 0 \\
        \vdots & \vdots & \ddots & \vdots & \vdots & \vdots \\
        0 & 0 & \cdots & C_{4} & 0 & 0 \\
        0 & 0 & \cdots & 0 & C_{2} & 0
    \end{pmatrix}.
\end{equation}
The terms containing the $\delta$-function type interactions are given by
\begin{equation}
\begin{split}
    H_{\mathrm{int}} 
    &= \left[
    g \op{\downarrow\downarrow} + g_{\uparrow\downarrow} 
    \left( \op{\mathrm{S}} + \op{\mathrm{A}} \right) + g \op{\uparrow\uparrow}
    \right]\delta(|x_{1}-x_{2}|) \\
    &= \frac{1}{\sqrt{2}}\left[
    g \op{\downarrow\downarrow} + g_{\uparrow\downarrow} 
    \left( \op{\mathrm{S}} + \op{\mathrm{A}} \right) + g \op{\uparrow\uparrow}
    \right]\delta(x_{-}).
\end{split}
\end{equation}
As the interaction does not affect the COM degree of freedom or the pseudo-spin degrees, the interaction energy matrix elements are given by 
\begin{equation}
\begin{split}
    \mel{n,2u,\downarrow\downarrow}{H_{\mathrm{int}}}{n,2\nu,\downarrow\downarrow} 
    &= \frac{g}{\sqrt{2}}\int dx_-\phi_{2u}(x_-)\phi_{2\nu}(x_-)\delta(x_-) \\
    &= \frac{g}{\sqrt{2}}\phi_{2u}(0)\phi_{2\nu}(0),
\end{split}
\end{equation}
\begin{equation}
    \mel{n,2u,\mathrm{S}}{H_{\mathrm{int}}}{n,2\nu,\mathrm{S}}
    = \frac{g_{\uparrow\downarrow}}{\sqrt{2}}\phi_{2u}(0)\phi_{2\nu}(0),
\end{equation}
\begin{equation}
    \mel{n,2u,\uparrow\uparrow}{H_{\mathrm{int}}}{n,2\nu,\uparrow\uparrow} 
    = \frac{g}{\sqrt{2}}\phi_{2u}(0)\phi_{2\nu}(0),
\end{equation}
\begin{equation}
    \mel{n,2u+1,\mathrm{A}}{H_{\mathrm{int}}}{n,2\nu+1,\mathrm{A}} 
    = \frac{g_{\uparrow\downarrow}}{\sqrt{2}}\phi_{2u+1}(0)\phi_{2\nu+1}(0) = 0,
\end{equation}
where $\phi_n(x)$ are the eigenstates of harmonic oscillator in the position representation, and where
\begin{equation} \label{eq:fuv}
\begin{split}
    \frac{1}{\sqrt{2}}\phi_{2u}(0)\phi_{2\nu}(0)
    &= \frac{1}{\sqrt{2\pi}}\left(-\frac{1}{2}\right)^{u+\nu}\frac{\sqrt{(2u)!(2\nu)!}}{u!\nu!} \\
    &\equiv f_{u,\nu} = f_{\nu,u}.
\end{split}
\end{equation}
Note that off-diagonal terms $f_{u,\nu}$ for $u<\nu$ are smaller than diagonal terms $f_{u,u}$, and off-diagonal terms $f_{u,\nu}$ disappear for increasing difference $\nu-u$ (see Fig.~\ref{fig:fuv}). 
The interaction term $H_{\mathrm{int}}$ can then be represented using $f_{u,\nu}$ as
\begin{equation} \label{eq:Hintapp}
    H_{\mathrm{int}} =
    \sum_{n=0}^{\infty}\sum_{u=0}^{\infty}\sum_{\nu=0}^{\infty}
    f_{u,\nu}
    \left(g\op{n,2u,\downarrow\downarrow}{n,2\nu,\downarrow\downarrow}
    +g_{\uparrow\downarrow}\op{n,2u,\mathrm{S}}{n,2\nu,\mathrm{S}}
    +g\op{n,2u,\uparrow\uparrow}{n,2\nu,\uparrow\uparrow}
    \right)
\end{equation}
and expressed in matrix form as
\begin{equation}
    H_{\mathrm{int}} =
    \begin{pmatrix}
        F_{0,0} & 0 & F_{0,1} & 0 & 0 & 0 & F_{0,2} & 0 & 0 & \cdots \\
        0 & F_{0,0} & 0 & 0 & F_{0,1} & 0 & 0 & 0 & 0 & \cdots
        \\
        F_{1,0} & 0 & F_{1,1} & 0 & 0 & 0 & F_{1,2} & 0 & 0 & \cdots
        \\
        0 & 0 & 0 & F_{0,0} & 0 & 0 & 0 & F_{0,1} & 0 & \cdots
        \\
        0 & F_{1,0} & 0 & 0 & F_{1,1} & 0 & 0 & 0 & 0 & \cdots
        \\
        0 & 0 & 0 & 0 & 0 & F_{0,0} & 0 & 0 & 0 & \cdots
        \\
        F_{2,0} & 0 & F_{2,1} & 0 & 0 & 0 & F_{2,2} & 0 & 0 & \cdots
        \\
        0 & 0 & 0 & F_{1,0} & 0 & 0 & 0 & F_{1,1} & 0 & \cdots
        \\
        0 & 0 & 0 & 0 & 0 & 0 & 0 & 0 & F_{0,0} & \cdots
        \\
        \vdots & \vdots & \vdots & \vdots & \vdots & \vdots & \vdots & \vdots & \vdots & \ddots
    \end{pmatrix},
\end{equation}
where each element is a four-by-four matrix such that
\begin{equation} \label{eq:matF}
\begin{split}
    F_{u,\nu} &=
    \begin{pmatrix}
        f_{u,\nu}g & 0 & 0 & 0 \\
        0 & f_{u,\nu} & 0 & 0 \\
        0 & 0 & f_{u,\nu} & 0 \\
        0 & 0 & 0 & 0
    \end{pmatrix} \\
    &= F_{\nu,u}.
\end{split}
\end{equation}
Similar to Eq.~(\ref{eq:Hsp_compact}) it can be written in a compact form as
\begin{equation}
    H_{\mathrm{int}} =
    \begin{pmatrix}
        \mathcal{F}_{0,0} & 0 & \mathcal{F}_{0,1} & 0 & \mathcal{F}_{0,2} & 0 & \cdots \\
        0 & \mathcal{F}_{0,0} & 0 & \mathcal{F}_{0,1} & 0 & \mathcal{F}_{0,2} & \cdots \\
        \mathcal{F}_{0,1}^{\dagger} & 0 & \mathcal{F}_{1,1} & 0 & \mathcal{F}_{1,2} & 0 & \cdots  \\
        0 & \mathcal{F}_{0,1}^{\dagger} & 0 & \mathcal{F}_{1,1} & 0 & \mathcal{F}_{1,2} & \cdots \\
        \mathcal{F}_{0,2}^{\dagger} & 0 & \mathcal{F}_{1,2}^{\dagger} & 0 & \mathcal{F}_{2,2} & 0 & \cdots \\
        0 & \mathcal{F}_{0,2}^{\dagger} & 0 & \mathcal{F}_{1,2}^{\dagger} & 0 & \mathcal{F}_{2,2} & \cdots \\
        \vdots & \vdots & \vdots & \vdots & \vdots & \vdots & \ddots 
    \end{pmatrix},
\end{equation}
where for $N\leq M$ the matrix elements are given by the four-by-four matrices in Eq.~(\ref{eq:matF}) and the additional $M-N$ columns are filled with zeros 
\begin{equation}
\mathcal{F}_{N,M} =
\left[
\begin{array}{cccccccccccccccccccc}
F_{N,M} & 0 & \cdots & 0 & 0 & 0 & \cdots & 0 \\
0 & F_{N-1,M-1} & \cdots & 0 & 0 & 0 & \cdots & 0 \\
\vdots & \vdots & \ddots & \vdots & \vdots & \vdots & \cdots & \vdots \\
0 & 0 & \cdots & F_{1,M-N+1} & 0 & 0 & \cdots & 0 \\
0 & 0 & 0 & 0 & F_{0,M-N} & \undermat{M-N}{0 & \cdots & 0} \\
\end{array}
\right].
\vspace{0.02\textwidth}
\end{equation}
The full Hamiltonian can then finally be expressed in compact form as
\begin{equation} \label{eq:Hfull_compact}
    H = 
    \begin{pmatrix}
        \mathcal{A}_0 + \mathcal{F}_{0,0} & \mathcal{B}_1 & \mathcal{C}_2 + \mathcal{F}_{0,1} & 0 & \mathcal{F}_{0,2} & 0 & \cdots \\
        \mathcal{B}_1^{\dagger} & \mathcal{A}_1 + \mathcal{F}_{0,0} & \mathcal{B}_2 & \mathcal{C}_2 + \mathcal{F}_{0,1} & 0 & \mathcal{F}_{0,2} & \cdots \\
        \mathcal{C}_2^{\dagger} + \mathcal{F}_{0,1}^{\dagger} & \mathcal{B}_2^{\dagger} & \mathcal{A}_2 + \mathcal{F}_{1,1} & \mathcal{B}_3 & \mathcal{C}_4 + \mathcal{F}_{1,2} & 0 & \cdots \\
        0 & \mathcal{C}_2^{\dagger} + \mathcal{F}_{0,1}^{\dagger} & \mathcal{B}_3^{\dagger} & \mathcal{A}_3 + \mathcal{F}_{1,1} & \mathcal{B}_4 & \mathcal{C}_4 + \mathcal{F}_{1,2} & \cdots \\
        \mathcal{F}_{0,2}^{\dagger} & 0 & \mathcal{C}_4^{\dagger} + \mathcal{F}_{1,2}^{\dagger} & \mathcal{B}_4^{\dagger} & \mathcal{A}_4 + \mathcal{F}_{2,2} & \mathcal{B}_5 & \cdots \\
        0 & \mathcal{F}_{0,2}^{\dagger} & 0 & \mathcal{C}_4^{\dagger} + \mathcal{F}_{1,2}^{\dagger} & \mathcal{B}_5^{\dagger} & \mathcal{A}_5 + \mathcal{F}_{2,2} & \cdots \\
        \vdots & \vdots & \vdots & \vdots & \vdots & \vdots & \ddots
    \end{pmatrix}.
\end{equation}

\section*{References}
\bibliographystyle{iopart-num}
\bibliography{bib_twosoc}

\end{document}